# Statistical Link Label Modeling for Sign Prediction: Smoothing Sparsity by Joining Local and Global Information

Amin Javari*, HongXiang Qiu*, Elham Barzegaran†, Mahdi Jalili‡, Kevin Chen-Chuan Chang*

\* Univesity of Illinois at Urbana-Champaign    † University of Lausanne    ‡ Royal Melbourne Institute of Technology
\* {javari2, hqiu6, kcchang}@illinois.edu    † elham.barzegaran@chuv.ch    ‡ mahdi.jalili@rmit.edu.au

*Abstract*—One of the major issues in signed networks is to use network structure to predict the missing sign of an edge. In this paper, we introduce a novel probabilistic approach for the sign prediction problem. The main characteristic of the proposed models is their ability to adapt to the sparsity level of an input network. The sparsity of networks is one of the major reasons for the poor performance of many link prediction algorithms, in general, and sign prediction algorithms, in particular. Building a model that has an ability to adapt to the sparsity of the data has not yet been considered in the previous related works. We suggest that there exists a dilemma between local and global structures and attempt to build sparsity adaptive models by resolving this dilemma. To this end, we propose probabilistic prediction models based on local and global structures and integrate them based on the concept of smoothing. The model relies more on the global structures when the sparsity increases, whereas it gives more weights to the information obtained from local structures for low levels of the sparsity. The proposed model is assessed on three real-world signed networks, and the experiments reveal its consistent superiority over the state of the art methods. As compared to the previous methods, the proposed model not only better handles the sparsity problem, but also has lower computational complexity and can be updated using real-time data streams.

## I. INTRODUCTION

Various types of interactions in diverse domains and systems can be modeled by complex networks. The examples of such systems include social networks, biological, transportation and information systems [2], [8]. Most of the studies about network modeling mainly consider the existence or absence of the connections between the nodes [18]; i.e., all relations between the nodes have the same concept and meaning. However, in some social networks, the nodes may have multiple types of relationships. A class of such networks can be modeled as networks with positive and negative links where a positive link means friendship (or trust) and negative link corresponds to enmity (or distrust) [10]. We may name Epinions, eBay, Wikipedia and Slashdot as web services from social network domain in which users may have different types of relations (i.e., they are signed networks) [9]. For instance, on eBay, a user can give trust or distrust rates for other active users. On Wikipedia, administrating users give positive or negative votes for the promotion of other users.

As much of the real-world is captured by complex networks with links characterizing the interactions of objects, predicting the labels of links becomes a crucial problem for exploiting networked information [13]. Indeed, in networks with multi-type of relations, not only the existence of the link between two nodes is of great importance, but also its label is as well. The problem of inferring labels of relations and in particular sign of a relationship has attracted attention in recent years under the topic: "sign prediction or link classification" [19]. Sign prediction has various applications in various domains. For example, recommender systems can benefit from efficient sign prediction algorithms [6]. Such algorithms can also be used to identify malicious users in signed networks [17].

Clearly, to solve the sign prediction problem, we seek to construct a model that is efficient as well as effective in terms of accuracy. Moreover, it is desirable to have a model that can incrementally adapt to changes in network structures because a growing number of emerging business, where high rate streams of detailed data are constantly generated, necessitate the need to build such models. Finally, it would be an interesting capability of a model to be able to perform on networks with more than two types of relations.

For sign prediction, we are facing the major challenge of the *sparsity* of information that plagues any complex real-world network [12]. Indeed, this is a key issue when deciding what information sources should be exploited for addressing the prediction task. In the sense of granularity, structures within a network can be viewed on two levels; local structures and global structures [15]. Roughly speaking, local structures are often described as structures at the level of the paths surrounding target edges while global structures refer to connectional patterns between communities of nodes.

One the one hand, some existing works solely rely on local information to predict the sign of the edge in question which makes them vulnerable to the sparsity problem. A representative example of such models is a supervised learning-based predictor proposed by Leskovec et al. [9]. The model, first extracts a set of features from the triangles involving the target edge; then, a learning method is employed to build a classifier to solve the sign prediction problem [9]. The proposed features in this work were derived based on two theories in signed networks, known as balance theory and status theory [1]. Both balance and status theories specify the patterns of connection at the level of triangles. However, robustness of the model against sparsity problem is quite limited as many nodes in networks do not share common neighbors [9]. To remedy the problem, Chiang et al. defined a new set of features by generalizing the notion of balance theory to $l$-cycles where a $l$-cycle is a path with length $l$ from a node to itself. In fact, by increasing $l$, the extracted features become less localized [11]. Although the model addresses the problem to some extent, still it is not able to capture global structure at the level of communities; moreover, the feature extraction process is computationally expensive, and the number of features exponentially increases with the length of cycles [15]. Very recently, a new line of research has aimed to solve the problem with the aid of embedding models [21], [22]. However, these models are not equipped to leverage the valuable information hidden in global structures because their proximity measures are not able to preserve global structures [20].

On the other hand, other existing works seek to use global structure to approach the problem which brings them robustness toward sparsity. It has been shown that if local structures in a signed network follow balance theory, it leads to a clusterability property of signed network stating that a complete network is weekly balanced if it can be partitioned into $K$ clusters in a way that all the edges between clusters are negative and all the edges within clusters are positive [12], [16]. With the help of this property, Hsieh et al. showed that signed networks exhibit a low-rank structure and formulated the sign prediction problem as a low-rank matrix completion problem [12], [13]. Having this formulation, available algorithms for matrix completion can be used for sign prediction. In another work, again based on clusterability property derived from balance theory, a prediction model was developed by finding the similarities of connectional patterns between clusters [16]. Naturally, cluster-based models are more robust against sparsity, and the obtained results from these works confirm that global structures may indeed have substantial merits for prediction purpose [12], [16]. However, such models are not able to extract information form local structures which can negatively affect their prediction accuracy.

In all, none of the previously introduced models fully leverage the rich information contained in both local and global structures. Besides this major issue, most of them have been designed specifically for signed networks as they have a basis in theories that are applicable only on signed network and extending such models for the general problem of link label prediction is non-trivial [9], [11], [16]. The other issue with many of the previous methods is that they cannot automatically adapt their prediction models to real-time data streams. In other words, to update those models based on the updated structure of input networks, prediction models should be built from scratch.

Towards addressing the prediction task effectively as well as solving the sparsity problem, as our key insight, we propose to holistically integrate and exploit both local and global information on a network. We suggest that indeed there exists a dilemma between these two information sources. Intuitively, local information specifically describes what is happening between target nodes, but the information obtained from local structures may end up being unreliable due to data sparseness. One the other hand, global structures tend to provide reliable information; however, naturally, such structures could not capture detailed information about target links. Indeed, it is challenging how to solve this dilemma between local and global structure optimally, i.e., when to use local structures and when to rely on global structures. A model that can solve this dilemma and exploit both information sources successfully not only would be able to address the sparsity issue but also can achieve promising results in terms of accuracy.

In this approach, we propose the concept of statistical link-label model, which can be defined as a probability distribution over the possible labels of the target edge, given the connections pointing out from the initiator node. The key characteristic of this model which enables us to exploit both local and global structures is that it reserves some parameters for the use of sophisticated smoothing techniques. For estimating those parameters, we develop a cluster based smoothing model through which we bridge the gap between local and global structures. This model considers global structure as a background knowledge and uses the background knowledge when sufficient local information is not available. Indeed, adaptivity of the proposed models comes from Dirichlet smoothing which enables us to make the balance between local and global structures. To build up the background knowledge, a novel clustering algorithm is developed. The proposed objective function used in this clustering algorithm is consistent with link label prediction task and can be optimized with the aid of optimization algorithms such as Gibbs sampling [7].

We show that the proposed model not only leverages both local and global structures but also addresses the aforementioned weaknesses of the counterpart models. We evaluate the models on three benchmark signed network datasets. Our results reveal that the proposed models consistently outperform previous sign prediction algorithms and demonstrate the high tolerance of the model toward the sparsity problem. We conjecture the main reason that accounts for the results is that the proposed link label model provides a principled way of exploring the local-global structures, which also makes the proposed predictors capable of adjusting themselves to the sparsity level of input networks. Moreover, the proposed model have lower computational complexity than the counterpart models. Also, the models can be updated incrementally with linear computational complexity on the total number of edges added to the input network. Indeed, the proposed model performs based on extracting some simple statistics from the network which accounts for its high efficiency. In addition, the model can be easily extended to perform on networks with more than two types of relations. Lastly, the structure of the proposed models have this advantage over the comparable models that it can easily accommodate contextual information and node attributes to provide more accurate predictions.

## II. Problem Formulation

In this section, we provide a formal definition of the signed prediction problem. A signed network is defined as a directed graph $G(V, E)$ with a link type mapping function $\varphi : E \rightarrow A$, in which all nodes $v \in V$ are of the same type, each link $e \in E$ belongs to one particular link type $\varphi(e) \in A$ where $A = \{+, -\}$. Suppose that over this network labels for some of the edges are missing. The problem is formulated as predicting the missing label for a given edge $e(u, v)$ from $u$ denoted as *initiator node* to $v$ denoted as *receiver node*, using the information on the labels from other edges. Note that link label prediction problem is the general case of the sign prediction problem where link labels may have more than two classes.

## III. SPARSITY AND DILEMMA BETWEEN LOCAL AND GLOBAL STRUCTURES

### A. Dilemma between local and global structures

Data sparsity is a prominent and critical issue when designing prediction models over networks. The performance of such models relies upon accurate and sufficient information being found in a network. However, it is known that abundantly sufficient data is not always available, in particular, on networks with high levels of sparsity. Imagine we are given a signed network, and the task is to predict the sign of the edge between two nodes contributing in quite a few links and sharing no common neighbors. Apparently, if we only rely on local information to determine the sign of the edge, we might make unreliable and inaccurate predictions due to lack of local information.

In general, it has been shown that for a large proportion of nodes in networks it is practically not possible to make reliable and accurate predictions by only relying on local structures. For example, in previously proposed path based models, it has been shown that prediction accuracy is considerably lower for the edges with lower values of embeddedness [9]. Embeddedness somewhat indicates the amount of local information available for prediction. We may draw this conclusion that, for the models based on local information, prediction cannot be performed accurately if local information about the target edge is not sufficient.

As a typical solution to the sparsity problem, the global structure of the network can be taken into account and patterns of connections between clusters may be utilized to approach the problem. Even on a network with high levels of sparsity, it is possible to assign a user to the cluster of users with similar connectional patterns and build a model based on the patterns of links between clusters. Intuitively, pattern extraction between clusters is less likely to encounter the sparsity problem because there always exist a substantial number of edges among clusters.

However, despite the fact that cluster-based models solve the sparsity problem, but they perform based on patterns that less specifically reflect the individual connections between target nodes. To our knowledge, most of the previously introduced models have chosen either local or global structures to build up their models. Indeed, models based on global structures may lose some rich and valuable local information while the predictors based on local structures might suffer from the sparsity problem.

This weakness of the previous models in fully capturing the local and global structures indicates a dilemma between reliability provided by global structure and specificity delivered by local data. A model that solves this dilemma in an intelligent manner not only would be able to tolerate against sparsity but also might outperform the counterpart models because it attempts to leverage more data.

### B. Solving the dilemma

In this paper, we propose a model that resolves the dilemma between global and local information; a model that can combine the information obtained from local and global structures in a principled way. This model can be viewed as a statistical model that enables us to compute the distribution of possible labels for the target edge given its *context* where context can be defined as the edges that their labels are somewhat related to that of the target edge. By computing such a distribution, it would be straightforward to infer the label of the edge in question.

The structure of the proposed model reserves some parameters for the use of sophisticated smoothing models which allows us to combine local and global structures in a principled way. By taking advantage of this characteristic, we propose the concept of cluster-based smoothing on networks. The main idea behind this concept is that the patterns extracted from the global structure of a network can be assumed as *background knowledge* on individual nodes. For estimating the parameters, we can more rely on this background knowledge as less reliable information can be gained from local structures. This concept of smoothing provides a rigorous theoretical foundation for integrating local and global analysis of networks and allows us to build prediction models that are tolerable and adaptive to sparsity.

It is worth mentioning that the structure of the proposed model and the cluster-based smoothing technique used in the model have some conceptual connections to statistical language models [14]. A language model can be described as a function to estimate the probability distribution for an upcoming word $w$ given its context, $P(w|context)$, where context is often defined as a sequence of previous or next words. Similar to our proposed models, one important characteristic of statistical language models such as n-gram model is their tolerance to data sparseness achieved by being able to employ smoothing techniques. For example, it has been shown that clustral structures of documents can be easily injected into such models through smoothing techniques [3].

## IV. LINK LABEL MODELING

The sign prediction problem can be approached by developing a probabilistic model to estimate the probability distribution over each of the classes of the target edge given the input graph. By comparing the obtained probabilities, the class with the highest probability may be defined as the final prediction:

$$\hat{l} = \arg\max_{l} P(\varphi(e(u_i, u_j)) = l | G(V, E)) \quad (1)$$

In this paper, we follow this direction and aim to approach the prediction model by proposing a probabilistic model named as *link label model* that estimates the probability distribution over possible labels for a link given its *context*, $P(\varphi(e(u_i, u_j)) = l | context)$.

If we view a link label model as a probabilistic classifier, the concept of context may be treated as the features of the classifier that should be extracted from the input graph $G(V, E)$. Now, the question is how to specify the context of the target edge. We suggest that the context of $e(u_i, u_j)$ can be defined as the set of the labels of connections initiated by $u_i$. Formally, context of $e(u_i, u_j)$ is denoted as $\varphi(E_i)$ where $E_i = \{e(u_i, u_x) | e(u_i, u_x) \in E\}$.

In language models, context of a word is typically defined as the window of words with a certain size before or after the word. We believe that analogous to the sequence of the words in a sentence, there exists a dependency between the labels of connections made by a user.

Now having the definition of context, the task is reduced to estimating the target distribution given the defined context. One very trivial approach to build the model is to generate all the possible contexts and build a model that has a parameter for every possible pair of the generated contexts and the target links. Clearly, with this approach, the number of parameters would be quite unmanageable.

This leads us to make some assumptions to build a simpler model with less parameters. As the simplest model, we may assume the labels of connections pointing out from a node are generated independently. Although such a model is simple, it makes an unrealistic assumption on the occurrence of labels of links. This assumption implies that when a user $u_i$ establishes a link toward $u_j$, the sign of the edge only depends on node $u_j$, i.e., in this model we have $P(\varphi(e(u_i, u_j)) = l | \varphi(E_i)) = P(\varphi(e(u_i, u_j)) = l)$. Therefore, we need to design a more complex model that is able to take the context of the target link into account.

Note that while, theoretically speaking, we would prefer to build a sophisticated model that can model occurrences of links more accurately, in reality, we face a tradeoff; as the complexity of model increases, the number of parameters increases as well. Therefore, the simplicity of the model is a factor that plays a crucial role in the final performance of the model.

In the next subsections, two link-label models based on local structures are presented. Then, we discuss how these two models can be modified to capture global structures. Finally, we introduce the concept of smoothing which allows us to extend the proposed models so that they can leverage both local and global structures.

### A. Link Label Modeling based on Local Structures

We propose two link label models based on local structures. The first model named as local structure based target link generator model (LTLGM) directly estimates the target distribution from the given context. However, the second model called as Local structure based Context Generator Model (LCGM) attempts to solve the reverse problem. That is, we can estimate the probability of generating the context given each possible label for the target edge and then assign a label to the target edge that is more likely to generate the context. Here, we introduce LTGM and LCGM is presented in the extended version of the paper.

*1) Target link label generator model:* As stated, LTGM directly estimates the target distribution from the given context. This model can also be viewed as a probabilistic discriminative classifier. Discriminative classifiers model the posterior $p(y|x)$ directly, or learn a direct map from inputs $x$ to class labels $y$.

However, it is not feasible to build a model through enumerating all the possible combinations of link labels generated by a user due to that such a model has a potentially infinite number of parameters. In order to tackle this issue, instead of capturing the dependency between the entire context and the target link, we can capture the dependencies between each element of the context and the target link. As such, the target probability can be calculated as follows:

$$P(\varphi(e(u_i, u_j)) = l | \varphi(E_i)) = \sum_{u_x \in H_{u_i}} \pi_x P(\varphi(e(u_i, u_j)) = l | e(u_i, u_x))), \quad (2)$$

where $H_{u_i}$ is the set of users that $u_i$ is pointing out to, i.e., the heads of the links initiated by $u_i$, and $\pi_x$ denotes the weight of the model associated with $u_x$. In our experiments, we set $\pi_x = \frac{1}{|E_i|}$. Also, $P(\varphi(e(u_i, u_j)) = l | e(u_i, u_x))))$ can be estimated using Maximum Likelihood Estimation (MLE) and based on local structures:

$$P_{local}(\varphi(e(u_i, u_j)) = l | e(u_i, u_x))) = \frac{|T_{u_j, l} \bigcap T_{u_x, \varphi(e(u_i, u_j))}|}{|T_{u_j} \bigcap T_{u_x, \varphi(e(u_i, u_j))}|}. \quad (3)$$

where $T_{u_x}$ is the set of nodes that have a connection toward $u_x$, i.e., the tails of the edges pointing to $u_x$, and $T_{u_x, \varphi(e(u_i, u_x))}$ is a subset of $T_{u_x}$ in which the labels of the edges are the same as $\varphi(e(u_i, u_x))$. In fact, in this equation we calculate the proportion of nodes that have links to both $u_x$ with label $\varphi(e(u_i, u_x))$ and $u_j$ with label $l$ to those that have links to both $u_x$ with label $\varphi(e(u_i, u_x))$ and $u_j$ with either positive or negative labels. Ideally, it is desirable to estimate dependency probabilities in respect to the target initiator node. That is, we can first identify the users similar to the initiator node $u_i$, and then estimate the dependency probability by only considering the connection of such users. However, this strategy may reduce the number of available statistics to estimate the target probability.

It is worth mentioning that the structure of the proposed model is conceptually related to the concept behind $k$-skip-$n$-gram models where a certain distance $k$ allows a total of $k$ or less skips to build the $n$-gram model [5].

*2) Context Generator Model:* Since our ultimate goal of estimating target distributions is to address the prediction problem, instead of solving the original problem, the reverse problem may be solved. That is, we can estimate the probability of generating the context given each possible label for the target edge and then assign a label to the target edge that is more likely to generate the context. Following this idea, we propose another link label model named as Local structures based Context Generator Model (LCGM).

From another point of view, we can think of the context generator model as a generative classifier. In general, generative classifiers learn a model of the joint probability of inputs $x$ and labels $y$ denoted as $p(x, y)$, and do predictions by employing Bayes rules to compute $p(y|x)$ and finally choosing the most likely label $y$.

Using Bayes theorem, the probability of the label $l$ for the target edge $e(u_i, u_j)$ conditioned to $\varphi(E_i)$ can be computed as follows:

$$P(\varphi(e(u_i, u_j)) = l | \varphi(E_i)) = \frac{P(\varphi(E_i) | \varphi(e(u_i, u_j)) = l) P(\varphi(e(u_i, u_j)) = l)}{P(\varphi(E_i))}, \quad (4)$$

where $P(\varphi(E_i) | \varphi(e(u_i, u_j)) = l)$ is the probability of observing $\varphi(E_i)$ conditioned to the fact that $\varphi(e(u_i, u_j)) = l$ and $P(\varphi(E_i))$ represents the probability of observing $\varphi(E_i)$. Since this is a classification task and $P(\varphi(E_i))$ is a constant for all of the classes, it can be removed from the prediction process. $P(\varphi(e(u_i, u_j)) = l)$ is our prior knowledge about class $l$ which can be set to a uniform distribution. Setting $\varphi(e(u_i, u_j)) = l$ to the uniform distribution, Equation 4 can be reduced to:

$$P(\varphi(e(u_i, u_j)) = l | \varphi(E_i)) \simeq P(\varphi(E_i) | \varphi(e(u_i, u_j)) = l). \quad (5)$$

Now the task is to compute $P(\varphi(E_i) | \varphi(e(u_i, u_j)) = l)$. Calculating this probability is not trivial, because the space of possible connections for each node is too vast. Therefore, we assume that the labels of the links within $E_i$ are independent given $\varphi(e(u_i, u_j)) = l$. This simplification reduces Equation 5 to:

$$P(\varphi(e(u_i, u_j)) = l | \varphi(E_i)) \simeq \prod_{u_x \in H_{u_i}} P(\varphi(e(u_i, u_x)) | \varphi(e(u_i, u_j)) = l), \quad (6)$$

where $P(\varphi(e(u_i, u_x)) | \varphi(e(u_i, u_j)) = l)$ is the parameter of the model which can be estimated using MLE and by exploiting local structures as follows:

$$P_{local}(\varphi(e(u_i, u_x)) | \varphi(e(u_i, u_j)) = l) = \frac{|T_{u_j, l} \bigcap T_{u_x, \varphi(e(u_i, u_j))}|}{|T_{u_x} \bigcap T_{u_j, l}|}, \quad (7)$$

### B. Link-Label Modeling Based on Global Structures

In the previous section, we introduced two models that only take local structures into account to estimate their parameters. However, as stated, parameter estimation based on local structures is quite vulnerable to sparsity problem. Indeed, despite the fact that we provided some robustness in our prediction models against sparsity by imposing a few assumptions when designing the structure of our models, but we find that still, we encounter sparsity issue when estimating the parameters. Obviously, the estimations made in Equations 3 and 7 are reliable if the number of available samples is sufficient, otherwise, the unreliability of estimation would lead to inaccurate predictions. However, there is no guarantee for the availableness of such sufficient number of samples. This motivates us to develop link-label models that exploit global structures.

Following this idea, we adapt the proposed models so that they can perform based on global structures. To this end, we incorporated the notion of clustering. By leveraging graph clustering algorithms, one can put users/nodes with similar connectional patterns in the same clusters whereas nodes with different patterns of connection are in different clusters. We suggest that the model's parameters can be estimated by analyzing connection at the level of clusters rather than individual nodes. The intuition behind this idea is that in the absence of sufficient data between nodes we can focus on the link between the clusters they belong to because nodes within the same cluster are expected to exhibit similar connectional patterns. Clearly, the way we partition the network into clusters plays a key role in our model. In the next section, we will present the proposed clustering algorithm which has been specifically designed for this task. Based on the proposed idea, given a set of clusters over the network the parameters of the model specified in Equation 3 can be estimated as follows:

$$P_{global}(\varphi(e(u_i, u_j)) = l | \varphi(e(u_i, u_x))) = \sum_{u_x \in E_i} \pi_x \frac{|T^{\Omega(u_i)}_{(\Omega(u_x), \varphi(e(u_i, u_x)))} \bigcap T^{\Omega(u_i)}_{(\Omega(u_j), l)}|}{|T^{\Omega(u_i)}_{(\Omega(u_x), \varphi(e(u_i, u_x)))} \bigcap T^{\Omega(u_i)}_{(\Omega(u_j))}|}, \quad (8)$$

where $\Omega$ is a cluster mapping function $\Omega : V \rightarrow C$ in which each node $v \in V$ belongs to one particular cluster $\Omega(v) \in C$ and $T^{\Omega(u_i)}_{\Omega(u_x), \varphi(e(u_i, u_x))}$ represents nodes belonging to $\Omega(u_i)$ that have links toward the nodes within $\Omega(u_x)$ with label $\varphi(e(u_i, u_x))$.

Note that, in order to estimate the probabilities based on local structures, the entire set of users in the graph are taken into account due to the sparsity problem. However, to estimate the parameters based on clusters we do not face sparsity, and hence the probability may be estimated with respect to the initiator node. As it can be seen in Equations 8, only nodes that belong to the same cluster as the initiator node are considered to estimate the model.

In the rest of the paper, we refer to this model as Global structure based Target Link label Generator Model (GTLGM).

In the same fashion, to build a context generator model that benefits from the global structures, the parameters specified in equation 7 can be estimated as follows:

$$P_{global}(\varphi(e(u_i, u_x)) | \varphi(e(u_i, u_j)) = l)) = \frac{|T^{\Omega(u_i)}_{\Omega(u_x), \varphi(e(u_i, u_x))} \bigcap T^{\Omega(u_i)}_{\Omega(u_j), l}|}{|T^{\Omega(u_i)}_{\Omega(u_x)} \bigcap T^{\Omega(u_i)}_{\Omega(u_j), l}|}, \quad (9)$$

We name the method constructed based on this estimator as Global structure based Context Generator Model (GCGM).

### C. Integrating Local and Global Structures

In previous sections, we introduced models that either rely on global or local structures. However, as stated, our main goal in this paper is to propose a model able to solve the dilemma between these two information sources. To achieve this goal, we employ the concept of smoothing on graphs which can effectively solve the problem. We believe that the notion of smoothing can be systematically applied to graph mining algorithms and may be viewed as a critical component of probabilistic graph mining models.

The major principle in smoothing methods is that a sparsely estimated conditional model can be smoothed using a more densely estimated but simpler model. We consider the estimators based on local structures as sparsely estimated models, and treat the models based on clusters as simpler models with more reliable estimations. Therefore, estimations based on individual nodes can be smoothed with the estimations from global structures to build models that benefit from both of the information sources. Various methods have been introduced for smoothing such as Dirichlet and Katz for different purposes over different domains [4]. Here we specifically focus on Dirichlet model and aim to adopt it to our prediction models. The reason for this is that Dirichlet smoothing is capable of regulating the estimation obtained from the estimators according to the reliability provided by the sparsely estimated model. By employing Dirichlete smoothing, the estimator of LTLGM can be smoothed through the one introduced for GTLGM. As such, we define a smoothing based estimator as follows:

$$P(\varphi(e(u_i, u_j)) = l | \varphi(e(u_i, u_x))) = \\ (1 - \lambda) P_{local}(\varphi(e(u_i, u_j)) = l) | e(u_i, u_x)) + \\ \lambda P_{global}(\varphi(e(u_i, u_j)) = l | \varphi(e(u_i, u_x))), \quad (10)$$

where $P_{local}$ and $P_{global}$ are presented in Equations 3 and 8 respectively, and $\lambda$ is defined as:

$$\lambda = 1 - \frac{|T_{u_x} \bigcap T_{u_j, l}|}{|T_{u_x} \bigcap T_{u_j, l}| + \mu}. \quad (11)$$

where $\mu$ is a parameter for Dirichlete smoothing that should be tuned. In fact, $\lambda$ indicates the reliability of the estimations made based on local structures. We name the model obtained based on this estimator as Smoothing based Target Label Generator Model (STLGM).

The same idea can be employed to build a context generator model by integrating local and global structures. Formally, we define a smoothing based estimator:

$$P(\varphi(e(u_i, u_x)) | \varphi(e(u_i, u_j)) = l) = \\ (1 - \lambda) P_{local}(\varphi(e(u_i, u_x)) | \varphi(e(u_i, u_j)) = l) + \\ \lambda P_{global}(\varphi(e(u_i, u_x)) | \varphi(e(u_i, u_j)) = l), \quad (12)$$

in which $P_{local}$ and $P_{global}$ are specified in Equations 7 and 9 respectively, and $\lambda'$ is defined as:

$$\lambda' = 1 - \frac{|T_{u_j} \bigcap T_{u_x, \varphi(e(u_i, u_x))}|}{|T_{u_j} \bigcap T_{u_x, \varphi(e(u_i, u_x))}| + \mu}. \quad (13)$$

The model from this estimator is referred as Smoothing based Context Generator model.

## V. Clustering

The way we do clustering is a crucial factor that decides the performance of the smoothing algorithm. For example, one may randomly partition the network and then derive the probabilities from those randomly generated clusters. By adding the prior model obtained from such method of clustering, we would add some noise to our prediction model that may lead to inaccurate predictions.

As stated, in our models based on global structure, we suggested that to extract connectional patterns between individual nodes, in the absence of enough local data, the connections between the clusters they belong to can be analyzed. This implies that nodes within the same cluster share the similar connectional patterns and can be considered as structurally *equivalent* units. Therefore, to build our clustering model first, we specify the notion of equivalence between two nodes.

Here, we suggest that two nodes are *equivalent* if they have identical types of connections to equivalent others. For example, imagine that $u_a$ is equivalent to $u_b$ and $u_c$ is equivalent to $u_d$, and there exist two edges $e(u_a, u_c)$ and $e(u_b, u_d)$. The proposed notion of equivalence imposes $\varphi(e(u_a, u_c))$ must be the same as $\varphi(e(u_b, u_d))$.

With this definition, if we partition the network into clusters of equivalent nodes, all of the edges between every pair of the clusters would have identical labels. Indeed, here clusters can be viewed as *roles* carried by nodes. Imagine there are $K$ latent roles in a network and every node within the network is assigned to one of them and the relation between two nodes is decided by the roles they hold. It means that for example in a signed network, all of the edges from the nodes with role A to nodes with role B must have the same label. Having such a structure over the input network, we can infer a node's connectional pattern by analyzing the link patterns of the node with the same role as the target node.

For partitioning the network based on our definition of equivalence, we should solve an optimization problem where the task is to assign nodes to clusters so that the edges from one cluster to another have identical labels.

Nevertheless, it must be mentioned that practically it is not possible to achieve such an ideal set of clusters over the input network which means that we need to relax our goal of clustering. Therefore, the **objective of clustering** is defined as partitioning the network, so that the majority of the links from one cluster to another have the same label. That is, suppose that an edge's label from one cluster to another is a random variable with a probability distribution over its possible values: positive and negative which means that for a network with $K$ clusters we would have $K^2$ random variables, one random variable for each pair of the clusters. We seek to partition the network in a way that the entropy of these random variables is minimum. We define our objective function as the weighted summation of the entropies of the random variables and the weight of each random variable is determined by the number of the edges between the clusters associated with the random variable. Formally, the objective function for a cluster set $g$ over a signed graph $G$ is defined as follows:

$$\phi(g) = -\sum_{c_i \in g} \sum_{c_j \in g} |E_{c_i, c_j}| \\ \sum_{k \in \{+, -\}} p_{c_i, c_j}(k) log(p_{c_i, c_j}(k)) \quad (14)$$

where $|E_{c_i, c_j}|$ is the size of the set of the edges from cluster $c_i$ to $c_j$ and $p_{c_i, c_j}(k)$ indicates the probability that an edge from cluster $c_i$ to cluster $c_j$ has label $k$. Using ML estimator, this probability can be computed as:

$$p_{c_i, c_j}(k) = \frac{|E_{c_i, c_j}(k)|}{|E_{c_i, c_j}|} \quad (15)$$

where $|E_{c_i, c_j}(k)|$ is the size of the set of the edges from cluster $c_i$ to $c_j$ with label $k$.

In order to do clustering based on the introduced objective function, a potentially difficult optimization problem must be solved. Indeed, our task is a special case of the general problem of finding the minimum of an objective function over a large combinatorial set. This type of problems is often amenable to Markov Chain Monte Carlo (MCMC) optimization, in which the model runs a Markov chain on the target combinatorial set and evaluates objective function at the successive states [7]. We employed an optimization model based on Gibbs sampling to construct our clustering algorithm. In the next subsection, we describe the optimization model.

*1) Clustering based on Gibbs sampling:* Our clustering algorithms can be described in two steps. First, we need to map the input combinatorial set to a graph that represents the neighborhood structure of the combinatorial set. Secondly, by making a random walk over the graph, one can reach a partition that fits the objective function under consideration [7]. In the following, we first define the neighborhood structure and then describe how a random walk can be made over this graph using Gibbs sampling algorithm.

Having our objective function, $\phi : P \rightarrow R+$, where $P = \{g_1, g_2, ..., g_m\}$ is the set of the all possible partitions on input graph and $m$ is the number of all possible partitions on the graph, we can define the neighborhood structure of $P$ as follows. Let's consider $GP$ as a connected, undirected graph with vertex set $P$ in which there is an edge between two vertices, $g_i$ and $g_j$, if and only if it is possible to get from partition $g_i$ to partition $g_j$ by moving exactly one of the $n$ objects in $g_i$ to a different cluster. In fact, the graph $GP$ represents a neighborhood structure on $P$; in which $g_i$ and $g_j$ are neighbors if and only if they have an edge in common. We denote the number of neighbors of the vertex $g$ in $GP$ as $d(g)$. Now the question is how to manage the random walk on graph $GP$.

As states, we employ a Gibbs sampler to make a random walk over $GP$. Suppose the current state of the Markov chain is the partition $g_i$, and $v_j$ a fixed object within the input graph. Let's suppose we remove the object $v_j$ from the partition $g_i$, and represent the obtained partition as $g_{i,-j}$ and define $S_{i,-j}$ as the set of partitions in $P$ that, when the object $v_j$ is removed are identical to $g_{i,-j}$. In our Gibbs sampling approach, to select a new state, we first evaluate the objective function for all partitions within $S_{i,-j}$ and then draw the new state from $S_{i,-j}$ based on the probabilities proportional to the objective function. In each iteration, the target object $v_j$ should be varied in order to cover all the objects within the network. There are two strategies to select the target object at each iteration: deterministic Gibbs sampler and random scan Gibbs sampler. In the deterministic version, $v_j$ is selected in a predetermined order while in the random scan version $v_j$ is selected randomly from $N$.

## VI. COMPUTATIONAL COMPLEXITY

In this section, we briefly discuss the computational complexity of the proposed prediction model. As stated in previous sections, in order to make predictions using our local structured based models, we first need to find the list of outgoing edges from target initiator node and then estimate the probabilities introduced in equations 3 and 7. Note that these parameters are estimated based on the number of nodes that have connections to both $u_x$ with label $l$ and $u_j$ with label $l'$. Those statistics of the input network can be computed in an offline manner and we don't need to wait to calculate them until the prediction time. Therefore, they can be stored in a matrix named as Node based Aggregation Matrix (NAM) with the size $|V|^2 \times 2 \times 2$ in which element $a_{m,n,l,l'}$ is the number of nodes that have a connection toward $u_m$ with label $l$ and a connection toward $u_n$ with label $l'$. NAM can be extracted from the network by one pass of scanning the edges and hence the computational complexity of building NAM is $O(E)$. Also, if we add $E'$ edges to the network, the cost of updating NAM would be $O(E')$.

To add the smoothing part to the basic prediction models, we should first partition the network and then estimate the parameters based on global structures as stated in equations 8 and 9. The computational complexity of clustering algorithm based on Gibbs sampling algorithm is $O(EI)$. According to our experiments, practically, the optimization algorithm converge after $10 \sim 20$ iterations on our benchmark networks.

To estimate the parameters based on global structures, again we can use statistics that could be calculated in an offline manner. Those statistics can be obtained from a Matrix named as Cluster based Aggregation Matrix (CAM) with the size $|C|^3 \times 2 \times 2$, in which $a'_{s,m,n,l,l'}$ is the number of nodes from cluster $C_s$ that have connections to cluster $C_m$ with label $l$ and connections toward cluster $C_n$ with label $l'$. CAM can be constructed by one pass through the edges within the network. Overall, the computational complexity of constructing NAG and CAG is $O(E)$ and the complexity of partitioning the network is $O(EI)$. As mentioned one advantage of the proposed models is that they can be updated incrementally based on a live stream of data with linear complexity on the number of edges added to the underlying network. Imagine a prediction model is constructed based on an input network and we aim to update the model incrementally using a live stream of data containing new edges added to the networks. To do so, the NAM and CAM graphs can be updated by adding up to the weights of the links corresponding to the newly added edges. Obviously, this could be done by one pass through the new edges. In other words, to update the model we do not need to build the model from scratch. However, this is not the case for most of the counterpart models. Also, we claimed that the proposed model is a generic model for link label prediction. In this paper, the model is introduced on the basis of the sign prediction problem. However, the model can be easily extended to perform on networks with multi-type of relations. Consider the equation 3 in which the label for an underlying edge is predicted by finding the probability of positive and negative labels for the edge given its context. To do label prediction for a network with $|K|$ possible labels, we simply need to calculate the probability of each of the $|K|$ possible labels given the context and finally select the label with the highest probability. Clearly, the clustering algorithm should also be modified accordingly for networks with multi-types of relations. Note that, the comparable sign prediction algorithms do not possess this flexibility.

## VII. EXPERIMENTS

In this section, we compare the performance of the proposed link label prediction model with those of the state of the art. The performance of the methods is assessed over three real-world datasets: Epinions, WikiElection, and Slashdot. We also investigate the validity of the idea of leveraging global structure to remedy the sparsity problem through our experiments.

*A. Test Methodology*

We evaluate the performance of the proposed algorithms using 10-fold cross validation as the testing algorithm. In 10-fold cross validation, in each fold, 10% of the original dataset is considered as test dataset and 90% as training dataset. Accuracy of classification is a common metrics to evaluate the predictions. However, in our datasets, the number of positive edges is considerably greater than the number of negative edges, and comparing prediction methods based on the accuracy of original test sets could be misleading under some circumstances. Thus, balanced accuracy is used in our experiments as the evaluation metric. In this method, the experiments are performed on the original test and training datasets and mean of the true positive rate of the predictions on two classes is reported (+1 and -1 for positive and negative signs, respectively).

*B. Datasets*

In order to assess the performance of proposed methods, we tested them on three signed networks: Slashdot, Epinions, and WikiElection which have frequently been used as benchmarks in different works [19]. Table 1 provides some statistics on these datasets.

Epinions is a web service about online product review in which users can express their votes on products. Also, in this platform users can give positive (trustworthy) or negative (untrustworthy) votes to other users regarding their reviews on products. The relations among users can be modeled by a directed signed network where users are nodes and their votes on each other form directed signed links of the network. Similar to Epinions, Slashdot dataset is obtained from a web service (technology news website) in which users can flag each other as friend or foe in order to indicate their approval or disapproval of their comments. WikiElection dataset is the network obtained from users' votes for elections of administrators in Wikipedia. In Wikipedia election, users may give positive or negative votes for the promotion of other users. Evidently, the meaning of links and their sign in WikiElection dataset differs from the other two datasets. However, all three systems can be modeled and generalized by signed networks, and the obtained network can be used to analyze the original systems.

Table I. DATASETS STATISTICS

|  | Nodes | Edges | +Edges | -Edges |
|---|---|---|---|---|
| Epinions | 119217 | 841200 | 85.0% | 15.0% |
| Slashdot | 82144 | 549202 | 77.4% | 22.6% |
| WikiElection | 7118 | 103747 | 78.7% | 21.2% |

*C. Reliability of parameter estimation for LTLGM and LCGM*

To predict the label of the link from an initiator node $u_i$ with $|E_i|$ outgoing links toward a receiver node $u_j$ using LTLGM and LCGM $|E_i|$ and $2|E_i|$ parameters should be estimated, respectively. As mentioned, the reliability of estimating these parameters is a crucial factor in the accuracy of the predictions made by these models which can be defined by the number of available samples for estimation. The more samples available for estimation the higher the reliability of the estimated parameters.

In this experiment, we investigate whether a sufficient number of samples are available for estimating the parameters of LTLGM and LCGM over our benchmark networks. We applied the basic models on the test sets obtained from WikiElection, Slashdot and Epinions datasets and measured the number of available samples for estimating the target parameters. Figure 1 demonstrates the percentage of the parameters estimated with less than $K$ samples over the three datasets as a function of $K$. As it can be seen from the figure, a large proportion of parameters for all three datasets should be estimated based on a few samples. For example, more than 40%, 60% and 80% of the parameters on WikiElection, Epinions and Slashdot datasets for LTLGM should be estimated with less than four samples. On Slashdot dataset, for more than 60% of parameters, even a single sample does not exist to estimate the parameters.

The problem is more severe for LCGM, as in LCGM we need to estimate two times more parameters than LTLGM. These results clearly show that how sparsity problem can negatively affect our local structure based prediction models and it necessitate a principled model to deal with sparsity problem.

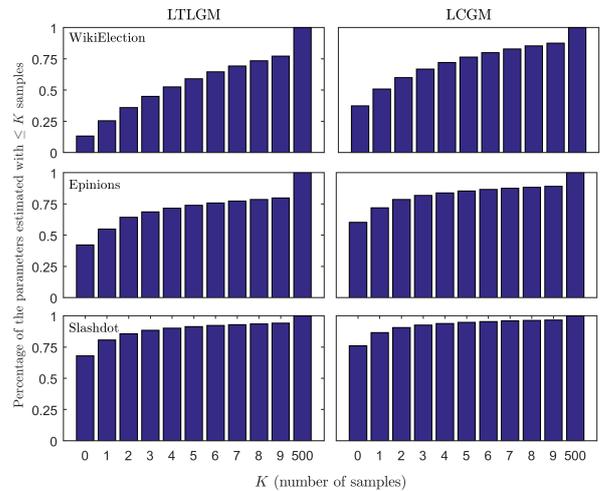

Figure 1. Percentage of the parameters of LTLGM and LCGM estimated with less than $K$ samples as a function of $K$ over WikiElection, Epinions and Slashdot datasets.

*D. The role of smoothing model on networks with different levels of sparsity*

In this experiment, we study the effectiveness of the proposed smoothing model on resolving sparsity problem for networks with different levels of sparsity.

To this end, first, we generated networks with various sparsity levels from our original networks by randomly removing edges from them. For example, to obtain a network with 10% density of the original Epinions datasets, 90% of the edges randomly selected and deleted from the network. Note that in this process, nodes were not removed. We applied our prediction models: LTLGM, LCGM, STLGM and SCMG models on the obtained datasets. Figures 2,3 and 4 represent the balanced accuracy of the models on networks with various sparsity levels generated from Epinions, Slashdot and WikiElection datasets respectively. As it can be seen, for all three datasets over different sparsity levels, STLGM and SCMG models substantially improve the accuracy of the local structure based models. This improvement is more notable on Slashdot and Epinions dataset compared with WikiElection dataset. This can be linked to the structure of the WikiElection dataset where the global structure of the network is not as informative as Epinions and Slashdot datasets. The other observation from this experiment is that as the sparsity of the networks increases, the outperformance of the models with smoothing technique over basic models becomes greater. This observation is quite consistent with our justification for the proposed model. As stated in previous sections, the prediction models based on local information are strongly vulnerable to sparsity problem while global structure has more tolerance toward sparsity. This is the reason that the idea of smoothing is more effective on sparser networks.

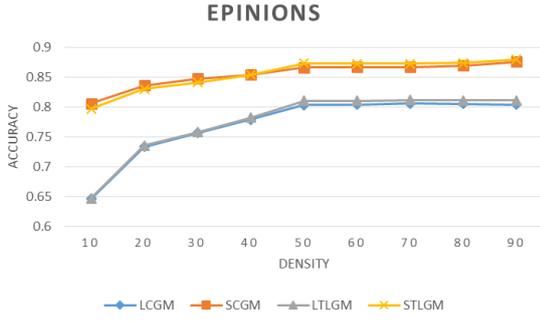

Figure 2. Accuracy of the proposed models: LTLGM, LCGM, STLGM, SCGM on networks with various sparsity levels obtained from Epinions dataset

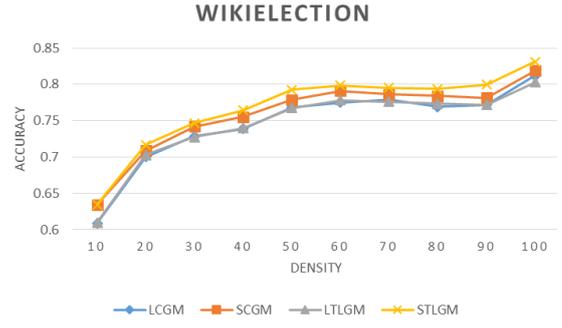

Figure 4. Accuracy of the proposed models: LTLGM, LCGM, STLGM and SCMG on networks with various sparsity levels obtained from WikiElection dataset

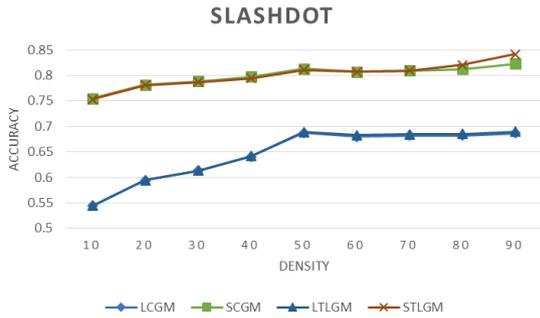

Figure 3. Accuracy of the proposed models: LTLGM, LCGM, STLGM and SCMG on networks with various sparsity levels obtained from Slashdot dataset

*1) Performance of the proposed models versus state-of-the-art sign prediction methods:* In this section, we compare the performance of the proposed predictors with six other predictors: those based on social balance theory (BT) and status theory (ST), two models based on machine learning framework: the one introduced by Leskovec et. al. (ML-23) [9] and the extended version proposed by Chiang et. al. (ML-HOC) [11], a model based on Matrix Factorization (MF) [15], a model based on user-based collaborative filtering (CF) [16] and a model based on node embedding (SiNE) [22]. Leskovec et. al. introduced 23 features for sign prediction and used Logistic regression as a classifier. Chiang et. al. extended this work by defining a new set of features extracted from higher order cycles. In our experiments, we extracted the features from cycles of order less than 6, as used in the original work of [11]. The MF-based predictor used in our experiments is the model introduced in [15]. Last, the embedding based predictor, denoted as SiNE, addresses the prediction task by learning vector representations for nodes through a deep learning framework where the objective function of the model is guided by balance theory [22].

Figure 5 compares the balanced accuracy of the methods on Epinions, Slashdot, and WikiElection datasets respectively. As it can be seen, the smoothing based models: STLGM and SCGM outperform both the models based on local structures: LTLGM and LCGM as well as those based on global structure: GTLGM and GCGM. Indeed, the accuracies of the individual models are even lower than the state of the art models, while integration of the individual models results in substantial improvements in terms of accuracy. These results shows that how the proposed model can effectively bridge the gap between local and global information sources and leverage them to build a powerful prediction model.

Also, the proposed smoothing based models presented better accuracy than other state-of-the-art sign prediction methods (i.e., those based on machine learning, social balance and status theories, matrix factorization, embedding and memory-based collaborative filtering). The superior performance of the model can be linked to the information sources used in our proposed model. While the state of the art models mainly rely on either local or global structures, our proposed model in a systematic way combines these two information sources using the idea of smoothing.

In all three datasets, predictors based on status and balance theories have the lowest accuracy values. It is beneficial to acknowledge that performances of the social theory based predictors depend on the meanings of the edges. For example in WikiElection dataset, users vote on each other based on the reliability of the trustee node. Users with higher reliability often receive more positive votes. Reliability of the trustor node can be interpreted as its status. It is the reason why the method based on status theory is a better predictor than the one based on balance theory in this dataset. In Slashdot and Epinions datasets, the votes represent the users' taste and preferences. Therefore, users' voting pattern could be better interpreted based on balance theory than status theory.

The features defined in path-based models also have a basis in these two theories. Similarly, the CF-based and the embedding based predictors also take advantage of the social theories. Therefore, these model these models can not be applied to address the link label prediction problem on networks with more than types of links. However, in our proposed models, we make no assumption about the meaning of the links, and in fact the models can be viewed as a predictor for the general problem of link label prediction.

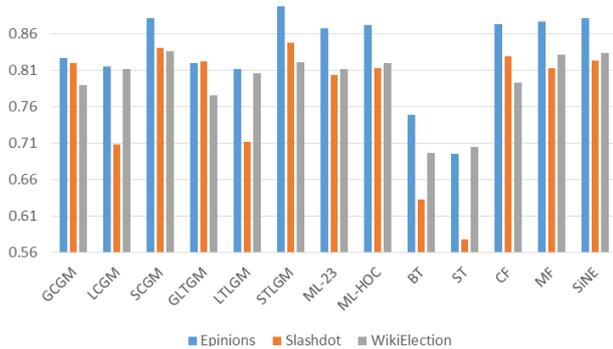

Figure 5. Accuracy of different predictors over Epinions, Slashdot and WikiElection datasets

By comparing the results in terms of consistency on different datasets, it is noticed that the proposed smoothing based deliver a consistent performance on all datasets, while this is not the case for other algorithms. We believe that this advantage is due to the fact that the proposed algorithm deals better with the sparsity problem. The proposed smoothing based models handle this problem by adapting themselves to the sparsity level of the networks. As the network becomes sparser, they give more weight for non-specific reliable approximations (which come from global structures) and as the reliability of the estimations from local information increases, they put more weight on local information.

## VIII. CONCLUSION

In this paper, we introduced a novel probabilistic approach for sign prediction named as link label modeling. Link label models allow us to find the distribution of possible labels for a target edge. We proposed two link label models based on local structures: LTLGM and LCGM, and two models based on global structures: GTLGM and GCGM. LTLGM and GTLGM models were combined to build another model named as STLGM using the idea of smoothing. Similarly, LCGM and GCGM were combined to build a model named as SCGM. It was shown that the hybrid models could adaptively combine the individual models which allows them to exploit both local and global structures in an intelligent manner. As the sparsity of the input network increases, they rely more on the models based on global structures while they give more weight to the models based on local structures when the sparsity decreases. We evaluated the proposed models on three real-world datasets, and the results showed that the proposed models outperform all previous methods. The proposed models can be generalized to solve the link classification problem. This feature of the model will be investigated in a separate work. Moreover, we believe that the performance of the proposed approach can be improved by employing hierarchical clustering algorithms. This idea will also be studied in a future work.